\documentclass[unsortedaddress,reprint,pra,aps,showpacs,floatfix]{revtex4-1}

\usepackage[draft]{hyperref}
\usepackage{amsmath}
\usepackage{graphicx}

\usepackage{amssymb,braket}
\newcommand{\ad}{a^{\dagger}}
\newcommand{\w}{\omega}
\newcommand{\tw}{\tilde{\omega}}
\newcommand{\al}{\alpha}
\newcommand{\bt}{\beta}
\newcommand{\gd}{\delta}
\newcommand{\gm}{\gamma}
\newcommand{\sx}{\sigma_x}
\newcommand{\sz}{\sigma_z}
\newcommand{\ds}{\hat{\rho}}
\newcommand{\ele}{\mathcal{L}}
\newcommand{\uu}{\mathcal{U}}

\begin{document}

\title{Decoherence and entanglement in tripartite systems: Cavity QED}

\author{E. Agudelo}
\affiliation{Instituto de F\'\i sica, Universidad de Antioquia, AA 1226, Medell\'\i n, Colombia}
\affiliation{Departamento de F\'\i sica, CP 702, Universidade Federal de Minas Gerais, 30123-970, Belo Horizonte, MG, Brazil}

\author{B. A. Rodriguez}
\affiliation{Instituto de F\'\i sica, Universidad de Antioquia, AA 1226, Medell\'\i n, Colombia}

\author{K. M. Fonseca-Romero}
\affiliation{Departamento F\'\i sica, Universidad Nacional de Colombia, Carrera 30 No. 45-03, Bogot\'a, Colombia}

\begin{abstract}
We propose an experimental setup, feasible with present day technology, involving two high-quality-factor cavities, one Ramsey zone and a two-level atom which interacts with them. The dynamics in the cavities is modeled by a dissipative dispersive Jaynes-Cummings model. Analytical expresions for the evolution superoperator in each part of the proposed setup and for the concurrence between the atom and the first cavity are given, and the pairwise entanglement of the subsystems is calculated numerically. We consider initial separable states given by the tensor product of a superposition of the two atomic relevant states and coherent states for the fields. We show that for realistic model parameters (including dissipation), when the dissipation rates are smaller than the dispersive frequencies (the ratio between the squared vacuum Rabi frequencies and the detunings), the final state of the cavity fields remains entangled.
\end{abstract}

\pacs{03.67.Bg,42.50.Pq,32.80.-t}

\maketitle

\section{Introduction}
Entanglement \cite{schrodinger1935,wootters1998,horodecki2009}, a characteristic feature of quantum mechanics, plays a key role in many quantum information tasks \cite{nielsen2000}. In order to implement those tasks, it is important to devise schemes for the creation of entangled states and to observe their dynamical characteristics in the presence of  enviromental effects. Cavity quantum electrodynamics (QED) systems, the cleanest open quantum systems known so far, are interesting from the point of view of transfering quantum information from
matter to radiation, from flying to static systems, and viceversa. Moreover, cavity QED setups \cite{haroche2001,walther2006} have been used in many fundamental experiments in quantum mechanics, such as the observation of decoherence \cite{brune1996}, the generation of atomic EPR pairs \cite{hagley1997} and the observation of quantum jumps \cite{gleyzes2007}.

This paper studies the entanglement dynamics under dissipation in a tripartite system, simple enough to allow for a quasi-analytical solution but rich enough to display a variety of dynamical characteristics,  which can be implemented experimentally with present day technology.
Figure \ref{fig:Scheme} shows the system, which is composed by several cavities crossed by a two-level atom. Similar proposals with multiple cavities, attempting to approach the Heisenberg limit for the estimation of an atomic phase shift \cite{vitali} or producing field-field entangled states \cite{davidovich}, can be found in the literature. In our system, the first and the last cavities are high quality cavities and the second one is a Ramsey zone. We calculate the degree of entanglement of the three pairs of subsystems, which are shown to behave as qubits for appropriate initial conditions, and discuss the role of the initial photon number and dissipation rate in the entanglement properties as measured by concurrence.

This paper has four sections: the first, section \ref{sec:model}, establishes the mathematical model that describes the proposed setup; the second (section \ref{sec:dynamics}) presents the evolution superoperator for the model; in the third section 
the state of the system when the atom traverses the first cavity and the atom-field concurrence are given; finally, section \ref{sec:cavity2} describes the behavior of pairwise concurrence obtained numerically by using typical experimental values.

\section{Mathematical model}
\label{sec:model}
\begin{figure}[h]
\includegraphics[width=0.48\textwidth]{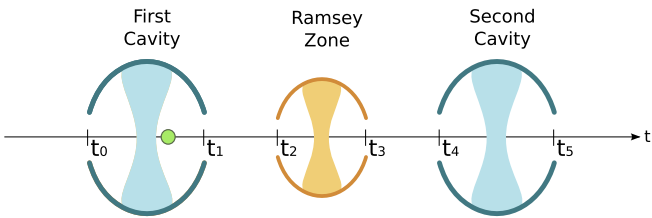}
\caption{\label{fig:Scheme}(Color online) Scheme of the proposed experimental setup.}
\end{figure}

We consider an apparatus composed of three cavities crossed by a two-level atom, as depicted in fig. \ref{fig:Scheme}. As it is clear from the picture, the atom undergoes several dynamical  processes: in the first step the atom interacts with a high quality dissipative cavity containing a coherent state, when the atom leaves this first cavity it will accumulate a phase due to free motion until the Ramsey zone is reached, where its state is rotated and, immediately thereafter a second free flight, the atom interacts with the final high-Q dissipative cavity prepared in a different coherent state. Although the cavities contain, in principle, infinite modes of the electromagnetic field, we consider only one mode per cavity field, almost resonant with the relevant atomic transition. It is important to realize that we are assuming that the center-of-mass coordinate of the atom is treated as if it were a classical variable, which is possible when the velocity of the atom is large enough not to be affected by the changes of the potential energy caused by the interaction with the fields. 

We assume that the whole setup is cooled to very low temperatures (of the order of 1K in this case) which allows us to model cavity field decay as interaction with a zero-temperature bath. We suppose that the atom-field detuning, in each cavity, is large enough to be in the dispersive regime. We can describe the electromagnetic field of the Ramsey zone as a classical one, in its interaction with the atom \cite{kim1999}. Taking into account these considerations we can write the equation of motion of $\ds$, the atom-fields state, as
\begin{equation}
\frac{d\ds(t)}{dt}=\frac{1}{i\hbar}\left[H,\ds(t)\right]+\ele_D(\ds(t)),
\end{equation}
where $\ele_D$, which models the dissipative effects, is written as
\begin{equation}
\sum_{i=1,2}\gm_i\left(2a_i\ds(t)\ad_i-\ds(t)\ad_i a_i -\ad_i a_i \ds(t)\right).
\end{equation}
The rate at which photons in the i-th cavity are lost, $2 \gm_i$, is proportional to the cavity frequency, $\tw_i$, and inversely proportional to the quality factor of the cavity, $\gm_i\propto\tw_i/Q_i$. Since we assume that the dissipative processes of the atom are much slower than those of the field, we ignore these processes in the description of the dynamics of the whole system. The Hamiltonian $H$ comprises two parts, which we designate by $H_0$ and $H_I$. The first part of the Hamiltonian, the free Hamiltonian $H_0$, is given by
\begin{align}
H_{0}= 
\frac{\hbar\w_a}{2}\sz 
+\sum_{i=1}^{2} \hbar\tw_{i}\left(\ad_i a_i+ \frac{1}{2}\right),
\end{align}
where $\w_a$ is the transition frequency between the relevant states of the atom, $\ket{g}$ (ground) and $\ket{e}$ (excited),  and $\sigma_z$ is the operator $\ket{e}\bra{e}-\ket{g}\bra{g}$. The creation $a_i^\dag$ and annihilation $a_i$ operators of the cavity fields, satisfy the usual bosonic commutation relations $[a_i,a_j^\dag]=\delta_{ij}$ and $[a_i,a_j]= 0 =[a_i^\dag,a_j^\dag]$. The second part of the Hamiltonian is the interaction Hamiltonian 
\begin{align}
H_I = H_{I1} W_{t_0,t_1}(t) + H_{R} W_{t_2,t_3}(t) + H_{I2} W_{t_4,t_5}(t),
\end{align}
where the function $W_{t_a,t_b}(t)$ is equal to $1$ if $t$ belongs to the interval $[t_a,t_b]$, and zero otherwise. Hence, the atom interacts with the first (second) cavity field in the interval of time $[t_0,t_1]$ ($[t_4,t_5]$) and with the Ramsey-zone field in the interval $[t_2,t_3]$. The coupling constant between the atom and the i-th cavity field is $\hbar \Omega_i$, where the vacuum Rabi frequency $\Omega_i$ is proportional to the product of the atomic dipole with the electric field having the root mean square amplitude of the vacuum in the cavity. The probability of detecting $n$ photons in the i-th cavity, $P_{n_i}=\braket{\ad_i a_i|\ds|\ad_i a_i}$, is usually employed to specify the conditions under which the dispersive approximation is valid. We define $N_i$, the relevant number of excitations in the i-th cavity, as the maximum number $n$, for which the probability $P_{n}$ is still important. If the detuning of the i-th cavity field, $\Delta_i = \omega_a -\tw_{i}$, satisfies the inequality $|\Delta_i|\gg \Omega_i \sqrt{N_i+1}$, we are in a region of parameters where the dispersive approximation is valid, and the atom-field interaction can be described by the hamiltonian \cite{savage1990}
\begin{equation}
\label{eq:ecuacionmaestra5}
H_{Ii}=\hbar\omega_i\left((\hat{a}^{\dag}_i\hat{a}_i+1)\ket{e}\bra{e}-\hat{a}^{\dag}_i\hat{a}_i\ket{g}\bra{g}\right), 
\end{equation}
with $\omega_i=\Omega_i^{2}/\Delta_i$. The Ramsey zone, which rotates the internal state of the atom, is modeled by the Hamiltonian
\begin{align}
H_{R} =e^{-\frac{i\w_R(t-t_2)}{2}\sz} H_{R0} e^{\frac{i\w_R(t-t_2)}{2}\sz} W_{t_2,t_3}(t),
\end{align}
where the Hamiltonian
\begin{align*}
H_{R0} = \left(\frac{\hbar \w_a}{2}\sz+\hbar \Omega_R|\xi_0|\sx\right),
\end{align*}
depends on  $\Omega_R$ the vacuum Rabi frequency of the Ramsey zone and $|\xi_0|$, the squared root of the mean number of photons of the coherent state sustained in the Ramsey zone. The operator $\sx$ is given by $\ket{e}\bra{g}+\ket{g}\bra{e}$. 

\section{Dynamics}
\label{sec:dynamics}
In order to solve the dynamics of the whole system in the interval $[t_0,t_5]$ it is convenient to find solutions in the intervals $[t_i,t_{i+1}]$, for $i=0,1,2,3,4$. For example, in the first interval we can write
\begin{align*}
\ds(t) = \uu_0(t,t_0) \ds(t_0) W_{t_0,t_1}(t),
\end{align*}
where the evolution superoperator $\uu_0(t,t_0)$ acts on $\ds(t_0)$, the initial state of the total system. The function $W$ stresses that the last equation is valid only in the interval $[t_0,t_1]$. The state of the system at the end of one interval is the initial state for the next interval. Thus, we can write, in general
\begin{align*}
\ds(t) = \uu_k(t,t_k) \ds(t_k) W_{t_k,t_{k+1}}(t),
\quad k=0,1,\ldots,4.
\end{align*}
Employing Lie algebraic techniques \cite{LieAlgebraicTechniques}, as shown in appendix \ref{sec:evolution}, we cast the evolution superoperator in the $[t_k,t_{k+1}]$ interval as
\begin{align}
\label{eq:evolutionsuperop}
\mathcal{U}_k(t,t_k) &
=U_k(t,t_k)\cdot U_k^\dag(t,t_k){\ }\mathcal{U}_{Dk}(t),
\end{align}
where $U_k(t,t_k)$ is the unitary operator which solves the Hamiltonian dynamics
\begin{align*}
i\hbar \frac{d}{dt} U_k(t,t_k) = 
\left( H_0 + H_I(t)\right) U_k(t,t_k),
\quad U_k(t_k,t_k)=I,
\end{align*}
in that time interval, and $\mathcal{U}_{Dk}(t)$ is a superoperator given below. We employ the dot-convention which establishes how (super)operators act on other (super)operators: the (super)operator to the right of the superoperator with dots will occupy the place of the dots. For example,
\begin{align*}
A \cdot B (\rho) = A \rho B,
\quad
A\cdot \left( B \cdot C \right) = A B \cdot C.
\end{align*}
The dissipative part of the evolution superoperator reads 
\begin{align}
\notag
\mathcal{U}_{Dk}(t)&=\prod_{i=1,2} e^{-\gamma_i(t-t_k) \left( \mathcal{M}_i+\mathcal{P}_i \right)} e^{\mathcal{F}_{ik} (t)\mathcal{J}_i}
\end{align}
where the operator $\mathcal{F}_{ik} (t)$ ($i=1, 2$; $k=0, 1, 2, 3, 4$)
\begin{align}
\mathcal{F}_{ik} (t)=\frac{2\gamma_i(1-e^{-2\gamma_i (t-t_k)-i \omega_{ik}(t-t_k) \left(\sz\cdot-\cdot \sz\right)})}
                 {2\gamma_i+i \omega_{ik} \left(\sz\cdot-\cdot \sz\right)} 
\end{align}
depends on the parameters $\w_{1k}=\w_{1} \gd_{k,0}$, $\w_{2k}=\w_{2} \gd_{k,4}$. 
The field operators $\mathcal{M}_i, \mathcal{P}_i$ and $\mathcal{J}_i$ are given explicitly by
\begin{align}
\label{eq:defMPJ}
\mathcal{M}_i = a_i^\dag a_i \cdot ,\quad
\mathcal{P}_i = \cdot a_i^\dag a_i,\quad \textrm{and}\quad
\mathcal{J}_i = a_i \cdot a_i^\dag .
\end{align}
The unitary contribution to the evolution superoperator can be cast as
\begin{align*}
U_k(t,t_k) = e^{-i(t-t_k)[H_0,\cdot]/\hbar} \tilde{U}_k(t,t_k)
\end{align*}
where $\tilde{U}_k(t,t_k)$ is the hamiltonian evolution operator in the interaction picture.
Hence, $\tilde{U}_1=I=\tilde{U}_3$ is the identity in the intervals of free flight; 
\begin{align*}
\tilde{U}_{k}(t,t_k) = e^{-i(t-t_k)[H_{I1/I2},\cdot]/\hbar}, 
\quad k=0,4,
\end{align*}
while the atom crosses the high-Q cavities and
\begin{align*}
\tilde{U}_2(t,t_2)=e^{-\frac{i\w_a}{2}(t-t_2)\sz}\exp\left[-i\Omega_R|\xi_0|(t-t_2)\sx\right],
\end{align*}
while the atom crosses the Ramsey zone, assumed to be resonant with the atomic transition.

\section{Entanglement and decoherence on the first cavity}
\label{sec:cavity1}
The initial state of the two-cavities-atom system that we consider, which is relatively easy to be
prepared with current experimental techniques, is a separable one: the atom in a superposition of its two relevant states and the fields in coherent states, $\ds(t_0)=\ket{\psi(t_0)}\bra{\psi(t_0)}$, where
\begin{align*}
\ket{\psi(t_0)} = \frac{1}{\sqrt{2}} \left(\ket{e}+e^{-i\phi/2}\ket{g}\right) \otimes \ket{\al} \otimes \ket{\bt},
\end{align*}
and $\ket{\al}$ ($\ket{\bt}$) is a coherent state for the first (second) cavity.
While the atom crosses the first cavity, the state of the system is
\begin{align}
\ds_1(t)&=\frac{1}{2}\left\{\ket{e,\al_1^e}\bra{e,\al_1^e}+x(t)\ket{e,\al_1^e}\bra{g,\al_1^g}\right.\\
&\left.+x^{*}(t)\ket{g,\al_1^g}\bra{e,\al_1^e}
+\ket{g,\al_1^g}\bra{g,\al_1^g} \right\}\otimes\ket{\bt_1}\bra{\bt_1},
\nonumber
\end{align}
where
\begin{align}
\label{eq:alphas}
\al_1^e & = \al e^{-i\tw_1 t-i\w_1 t -\gm_1 t}\nonumber\\
\al_1^g & = \al e^{-i\tw_1 t+i\w_1 t -\gm_1 t}\\
\bt_1   & = \bt e^{-i\tw_2 t -\gm_2 t}\nonumber
\end{align}
and
\begin{gather}
x(t)=\exp\left[-i\frac{\phi}{2}+|\al|^2 f_x(t)-i(\w_1+\w_a)t\right],\\ \nonumber
f_x(t)= \frac{\gm_1}{\gm_1+i\w_1}(1-e^{-2(\gm_1+i\w_1)t})-(1-e^{-2\gm_1t}).
\end{gather}
We have chosen $t_0=0.$

\begin{figure}[th]
\includegraphics[width=0.48\textwidth]{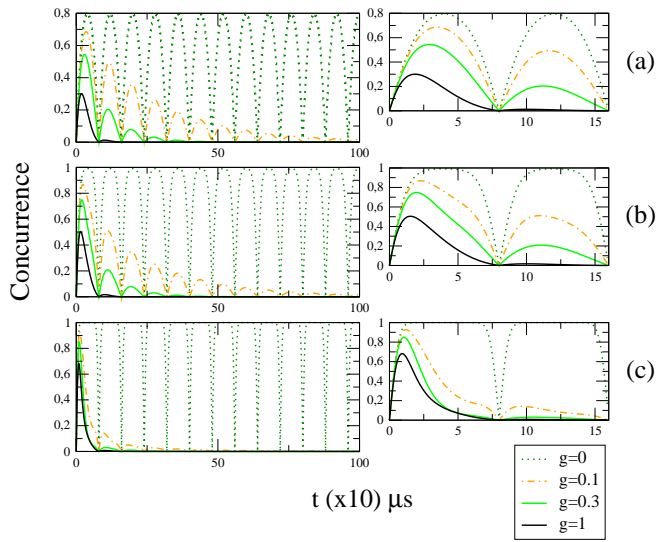}
\caption[Atom-first cavity field concurrence.]{(Color online) Atom-first cavity field concurrence while the atom crosses the first cavity, as a function of the initial mean number of photons, (a) $\al=$0.5, (b) $\al=$1, (c) $\al=$2 and for different values of the dissipation rate. In the second column a blow-up of the first two-oscillations is shown. For $t_0=0$ and $t_1=1000$ $\mu s$.}
\label{fig:a1}
\end{figure}

Since the atom-first cavity field density operator belongs to a $\mathbb{C}^2 \otimes\mathbb{C}^2$ subspace of the total
available Hilbert space, it corresponds to a couple of qubits, whose entanglement can be quantified by means of Wootters concurrence \cite{wootters1998}. In our case
\begin{align}
C(\ds_1)=|x(t)| \sqrt{1-|\braket{\al_1^g|\al_1^e}|^2}.
\label{eq:conc}
\end{align}
Clearly, one of the factors on which concurrence depends, is a function of the nonorthogonality of the states $\ket{\al_1^e}$ and 
 $\ket{\al_1^g}$, the more different the states the larger the concurrence. Although the other presents a more complex dependence, its main contribution is of the form
\begin{align*}
e^{-|\al|^2(1-e^{-2\gm_1t})},
\end{align*}
that is, it principally depends on the initial number of photons and the product of the decoherence rate and the time spent in the cavity. In fig. \ref{fig:a1} we show the time evolution of the concurrence between the atom and the first cavity for different values of initial mean number of photons and different dissipation rates, the time interval was chosen between $t_0=0$ and $t_1=1000$ $\mu s$. As the initial average photon number is increased, the maximum entanglement {lasts} for longer times in a {lossless} cavity. However, at the same time, for a lossy cavity, at larger values of $\al$, atom and cavity field separate faster.

\begin{figure}[th]
\includegraphics[width=0.48\textwidth]{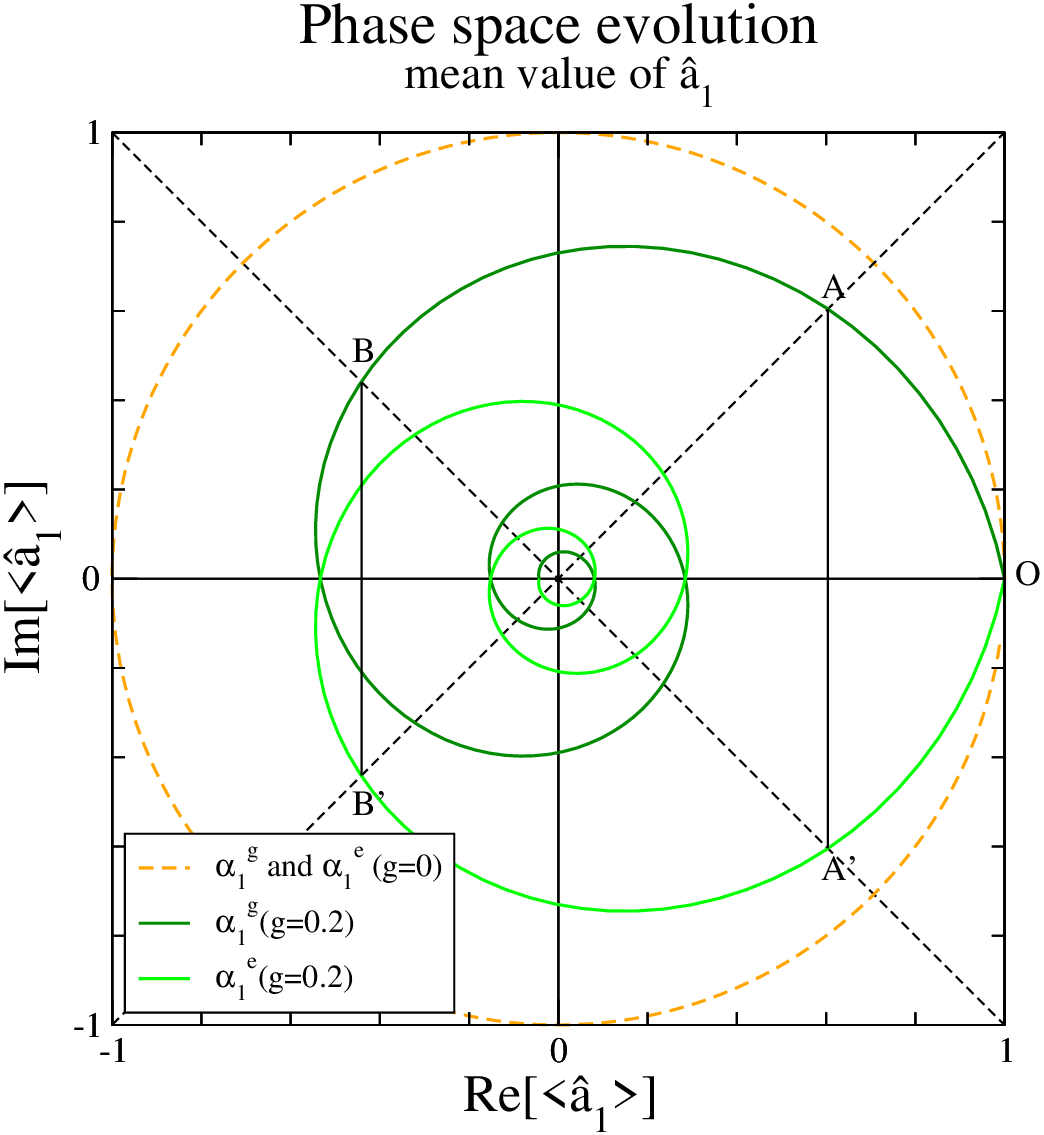}
\caption{\label{fig:phasespace}(Color online) Phase-space evolution of $\al^e_1$ and $\al^e_1$ in a rotating frame with frequency $\tw_1$. The initial mean number of photons is $\al=1$. Dashed line shows the dissipationless case and solid lines represents the dissipation one with $\gm_1=1.25$ $kHz$ and dispersive frequency $\w_1=6.25$ $kHz$. Vertical distance of the trajectories at same time is proportional to entanglement. The line $AA'$($BB'$)  corresponds to a time of larger (smaller) entanglement.}
\end{figure}

A graphic way to see this is by looking at the phase space representation of the field \cite{Schleich}. The physics of this part of the process seems easy to understand: without dissipation, the key element for entanglement, the interaction, is proportional to the field intensity. However, for coherent states, the role of the interaction is to split the field into a component correlated to the excited atomic state whose label rotates at angular velocity $\tw_1+\w_1$ and a component correlated to the ground atomic state whose label rotates at angular velocity $\tw_1-\w_1$ (See eq. (\ref{eq:alphas})). It is easier to concentrate on the angular velocity differences, and to plot the labels in a phase space picture (fig. \ref{fig:phasespace}). It is clear that every $2\pi/(2\w_1)$ seconds both labels coincide, hence the state becomes separable. This process remains unaffected in the presence of decoherence. From eq. (\ref{eq:conc}) it is clear that entanglement decreases as the overlap increases, or, in other words, entanglement decreases with the phase space distance between the field states correlated with the excited and ground states, distance which is given by the length of the vertical line that join those states. 
For larger mean number of photons, the phase space distance is large most of the time, except near the separation times; for small mean number of photons this distance never grows appreciable, neither does the entanglement.

In the presence of dissipation the characteristic time for decoherence is inversely proportional to the average photon number. We have already pointed out that a factor which decays with the characteristic time of decoherence enters in the expression for concurrence. In fig. \ref{fig:phasespace} one observes that the part of the field entangled with $\ket{e}$ ($\ket{g}$) rotates clockwise (counterclockwise) and its radius shrinks with dissipation. The assymetry of the entanglement curves shown in the fig. \ref{fig:a1} can be understood in terms of the phase space distance, as was discussed above, and marked on fig. \ref{fig:phasespace} with vertical lines.

As the atom crosses the second cavity, the expressions for the density operator calculated analytically become much more cumbersome and are omitted. Therefore, concurrence is calculated only numerically. From here on we make a more numerical approach. Taking into account typical values of current experiments  \cite{brune1996} we consider the following fixed parameters: the angular dispersive frequencies $\w_1=\w_2=6.25$ $kHz$, the atomic transition frequency and the Ramsey zone frequency $\w_a=\w_R= 51.1\times 10^6$ $kHz$, the detuning $|\Delta_1|=|\Delta_2|= 100$ $kHz$ and Rabi frequencies $\Omega_1=\Omega_2=25$ $kHz$. The flight times through the cavities are also fixed: the atom spents 30 $\mu s$ in each cavity and 10 $\mu s$ between cavities. The dissipation rates for the cavities $\gm_1$ and $\gm_2$ are varied according to $\gm_1=g\w_1$ and $\gm_2=q\w_2$, where $g,q=0,\ 0.05,\ 0.5,\ 1$. In order not to break the dispersive approximation, we can not increase too much the mean number of photons. To estimate the validity of the dispersive approximation we use $|\Delta_1|\gg \Omega \sqrt{|\al|^2+1}$. For the values of the parameters chosen above, we are constrained to coherent states satisfying $|\al|^2\sim \bar{n}\ll 15.$ We thus choose, for the mean number of photons, the values $\al,\bt=0.5, 1, 2$.

\section{Entanglement and decoherence on the second cavity}
\label{sec:cavity2}
	
\begin{figure}[t]
\includegraphics*[viewport=0 0 680 280,scale=0.48]{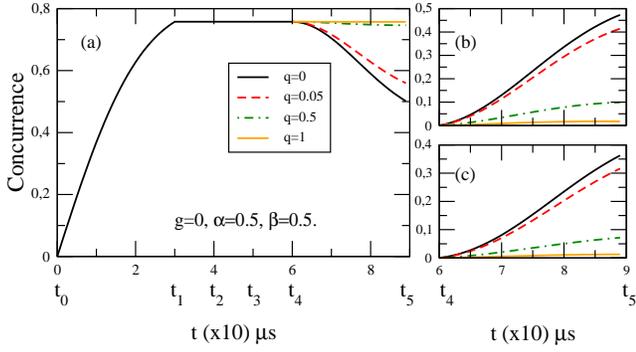}
\caption{\label{fig:ideal1}(Color online) Atom-field1 (a) atom-field2 (b) field1-field2 (c) concurrence for ideal first cavity and different dissipation rates for the second one.}
\end{figure}

We begin by considering the first cavity to be ideal and the second to range between ideal (q=0) and strongly dissipative (q=1). In fig. \ref{fig:ideal1} we plot the concurrences for all three pairs of subsystems. For ideal cavities, and initial coherent states with $\al=0.5=\bt$, the final state contains quantum correlations between all pair of subsystems. If the second cavity is strongly dissipative only the atom and the first field show relevant quantum correlations.

\begin{figure}[h]
\includegraphics*[viewport=0 0 680 280,scale=0.48]{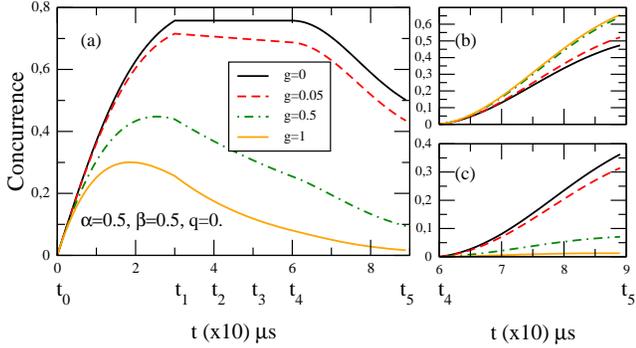}
\caption{\label{fig:ideal2}(Color online) Atom-field1 (a) atom-field2 (b) field1-field2 (c) concurrence for ideal second cavity and different dissipation rates for the first one.}
\end{figure}

If we consider the opposite case, that of a ideal second cavity, and first cavity ranging between ideal and strongly dissipative, entanglement behavior is complementary to the previous case (fig. \ref{fig:ideal2}): only the atom-second cavity field are important for a strongly dissipative first cavity.

\begin{figure}[t]
\includegraphics[width=0.48\textwidth]{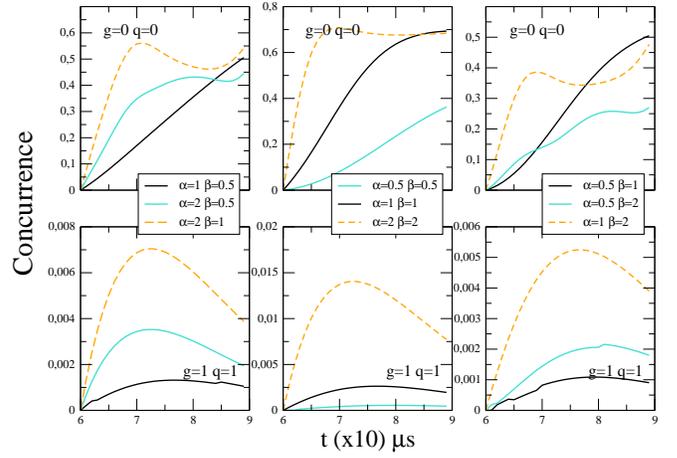}
\caption{\label{fig:transf}(Color online) Field1-Field2 concurrence. Maximum values are obtained for (nearly) ideal cavities and minimum values for strongly dissipative cavities.}
\end{figure}

We conclude that the most interesting physical effects from the point of view of multipartite entanglement happen for decoherence rates of about one order of magnitude smaller than the dispersive frequency. This conclusion holds even if we change the mean number of photons in the cavities, as shown in fig. \ref{fig:transf}. All these results shows that the local independent interactions between the atom and the cavity fields has as consequence the generation of non-local field-field entanglement.

\begin{figure}[h]
\includegraphics*[viewport=0 0 680 510,scale=0.35]{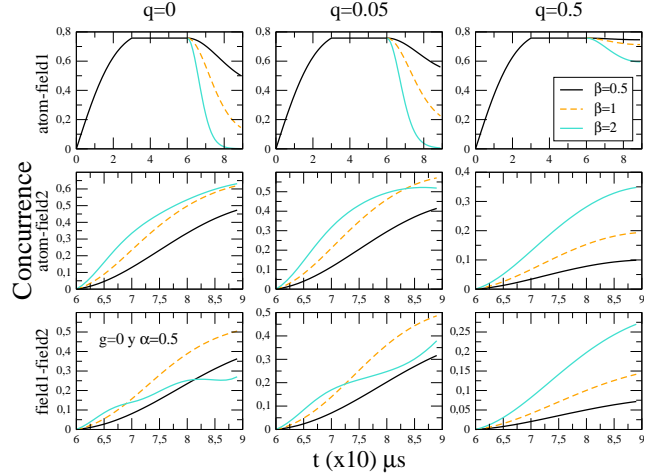}
\caption{\label{fig:allb}(Color online) Atom-field1, atom-field2 and field1-field2 concurrence for an ideal first cavity, $\al=0.5$ and different values of $q$ and $\bt$.}
\end{figure}

As noticed before, increasing the mean photon number can lead to smaller atom-cavity (quantum) correlations (see fig. \ref{fig:allb}). In fact, atom-first field concurrence decreases as the initial mean number of photons in second cavity increases from one to four. Also, if we begin with coherent states with mean photon numbers larger than one, even for not so large values of the decoherence rate, we observe sudden death of the atom-first field entanglement (see fig. \ref{fig:allb}, first row). As shown in figures \ref{fig:ideal1}-\ref{fig:ideal2} and \ref{fig:allb}, the entanglement between all pairs of subsystems can not increase simultaneously, suggesting some kind of \emph{entanglement conservation law}, related to the Coffman-Kundu-Wootters (CKW) monogamy inequality \cite{Coffman2000}.

It is interesting to notice that as the initial mean number of photons increase so does the field-field entanglement, for (nearly) ideal cavities, and short times. As time goes on, this rule does not hold anymore, and there seems to be an optimal value of the mean number of photons to produce the higher field-field entanglement for a given interaction time (see fig. \ref{fig:allb}, for g=0 and q=0, 0.5).  


In summary, we have considered a realistic cavity QED setup, assumed to be operated on the dispersive regime, and have solved analytically its dynamics, in the sense of finding the evolution superoperator. We have shown that for typical experimental values of the cavity dissipation rates, smaller than the dispersive frequencies, and times of the order of 100 $\mu s$, not only the final state generally exhibits atom-field and field-field entanglement, but that the effect of the dissipation is not trivial, in the sense that the pairwise concurrences are not monotonic functions of the dissipation rates.

\section*{Acknowledgments}
The authors thanks  H. Vinck, M. C. Nemes and J. G. Gon\c{c}alves de Oliveira
Junior, for very useful discussions. K.M.F.R. acknowledge partial support to
DIB-UNAL.

\appendix*

\section{Evolution superoperator}
\label{sec:evolution}

It is easy to show that the evolution superoperator $\mathcal{U}_k(t,t_k)$ in the interval $[t_k,t_{k+1}]$, $k=1,2,3$, given by
\begin{align*}
T \exp 
\int_{t_k}^t\!\!\! d\tau \left(\mathcal{L}_D+\frac{1}{i\hbar}[H,\cdot]+\frac{\delta_{k,2}}{i\hbar}[H_R(\tau),\cdot] \right),
\end{align*}
where $T$ is the time-ordering operator, 
can be written as \eqref{eq:evolutionsuperop} due to the following commutation relations
\begin{align*}
[H_0 \cdot, \mathcal{L}_D] & = 0 = [\cdot H_0, \mathcal{L}_D],\\
[H_R (t) \cdot, \mathcal{L}_D] & = 0 = [\cdot H_R (t), \mathcal{L}_D].
\end{align*}
In the intervals $[t_0,t_1]$ and $[t_4,t_5]$ the evolution superoperator
\begin{align*}
\mathcal{U}_k(t,t_k) = e^{\mathcal{L}(t-t_k)}, \quad k=0,4,
\end{align*}
can be written as
\begin{align}
\label{eq:app1}
\mathcal{U}(t,t_k) = e^{\mathcal{L}_{C}(t-t_k)} e^{\mathcal{L}_{N}(t-t_k)},
\end{align}
with
\begin{align*}
\mathcal{L}_{C} = \frac{1}{i\hbar}[H_A+H_F +\sum_i \omega_{ik} [a_i^\dag a_i \frac{\sz\cdot+\cdot \sz}{2} +\ket{e}\bra{e},\cdot],\\
\mathcal{L}_{N}  = \sum_i 
\left( 
2\gamma_i \mathcal{J}_i 
-\frac{2\gamma_i+i \omega_{ik} \left(\sz\cdot-\cdot \sz\right)}{2} 
\left( \mathcal{M}_i+\mathcal{P}_i \right)
\right),
\end{align*}
where $\w_{1k}=\w_{1} \gd_{k,0}$, $\w_{2k}=\w_{2} \gd_{k,4}$ and the operators $\mathcal{M}_i,\mathcal{P}_i, \mathcal{J}_i$ are defined in \eqref{eq:defMPJ}. Taking into account that the atomic operator commute with the field operators, we write
\begin{align*}
e^{\mathcal{L}_{N}(t-t_k)}  & = \prod_i e^{\alpha_{ik} \left( \mathcal{M}_i+\mathcal{P}_i \right)+\beta_{ik}\mathcal{J}_i}\\
&= \prod_i e^{\alpha_{ik} \left( \mathcal{M}_i+\mathcal{P}_i \right)}
e^{\frac{\beta_{ik}}{2\alpha_{ik}}(\exp(2\alpha_{ik})-1)\mathcal{J}_i}
\end{align*}
where the commutation relations
\begin{align*}
[ \mathcal{M}_i+\mathcal{P}_i, \mathcal{J}_j] = -2\mathcal{J}_j\delta_{ij},
\quad i,j=1,2,
\end{align*}
and disentangling operators techniques \cite{LieAlgebraicTechniques}
were used. It is clear that the coefficients $\al_{ik}$ contain a term proportional to $\gamma_i$ and one proportional to $\w_{ik}$, which can be separated. We can write therefore
\begin{align}
\label{eq:app2}
e^{\mathcal{L}_{N}(t-t_k)}  
&= \prod_i e^{\alpha_{ik}(\gamma_i=0) \left( \mathcal{M}_i+\mathcal{P}_i \right)}\\
&  \quad e^{\alpha_{ik}(\w_{ik}=0) \left( \mathcal{M}_i+\mathcal{P}_i \right)}
e^{\frac{\beta_{ik}}{2\alpha_{ik}}(\exp(2\alpha_{ik})-1)\mathcal{J}_i}.\nonumber
\end{align}
Grouping the first term of \eqref{eq:app1} with the terms above for zero dissipation ($\gamma_i=0$) we recover the hamiltonian dynamics. On the other hand the last line of \eqref{eq:app2} corresponds to the purely dissipative dynamics (with zero Hamiltonian). After employing the explicit expressions for $\al_{ik}, \bt_{ik}$, the evolution superoperator can be written as \eqref{eq:evolutionsuperop}.

\end{document}